\documentclass[prl,floatfix,twocolumn,showpacs,amsmath,amssymb]{revtex4}
\usepackage{graphicx}
\usepackage{dcolumn}
\usepackage{bm}
\usepackage{epsfig,psfrag,amsmath,amssymb,float}
\input{epsf}
\begin{document}

\title{Doped coupled frustrated spin-$\frac{1}{2}$ chains
with four-spin exchange}

\author{Nicolas Laflorencie and Didier Poilblanc}

\affiliation{Laboratoire de Physique Th\'eorique CNRS-FRE2603,, Universit\'e
Paul Sabatier, F-31062 Toulouse, France}

\date{\today}

\begin{abstract}

The role of various magnetic inter-chain couplings is investigated
by numerical methods in doped frustrated quantum spin chains. A
non-magnetic dopant introduced in a gapped spin chain releases a
free spin-$\frac{1}{2}$ soliton. The formation of a local magnetic
moment is analyzed in term of soliton confinement. A four-spin
coupling which might originate from cyclic exchange is shown to
produce such a confinement. Dopants on different chains experience an effective
space-extended non-frustrating pairwise spin interaction.

\end{abstract}

\maketitle

Strong interest for quasi-one-dimensional frustrated spin chains
has been revived since the discovery in 1993 of the first
inorganic spin-Peierls (SP) compound CuGeO$_3$ which exhibits a
transition~\cite{Hase93} towards a gapped dimerised ground state
(GS). Some vanadate oxides such as LiV$_2$O$_5$ are also excellent
realizations of weakly interacting frustrated spin-1/2
chains~\cite{LiV2O5}. Whether magneto-elastic couplings play a
role at low temperature in LiV$_2$O$_5$ is not yet clear although
it has been investigated theoretically~\cite{Becca2002}. Replacing
a spin-$\frac{1}{2}$ in a {\it spontaneously} dimerised (isolated)
spin chain by a non magnetic dopant (described as an inert site)
liberates a free spin $\frac{1}{2}$, named a soliton, which do not
bind to the dopant~\cite{Sorensen98}. The soliton can be depicted
as a single unpaired spin (domain) separating two dimer
configurations~\cite{Sorensen98}. The physical picture is
completely different when a {\it static} bond dimerisation exists
and produces an attractive potential between the soliton and the
dopant~\cite{Sorensen98,Nakamura99} and consequently leads, under
doping, to the formation of local magnetic
moments~\cite{Sorensen98,Normand2002} as well as a rapid
suppression of the spin gap~\cite{Martins96}. However, a coupling
to a purely one-dimensional (1D) adiabatic lattice~\cite{Hansen99}
does not produce confinement in contrast to more realistic models
including an elastic inter-chain coupling (to mimic 2D or 3D
lattices)~\cite{Hansen99,Dobry98}. Doping quasi-1D dimerised spin
chains is realized experimentally by substituting a small fraction
of Copper atoms by Zinc (or Magnesium) atoms in
CuGeO$_3$~\cite{Zn_CuGeO}.
\begin{figure}
\begin{center}
\psfrag{J1}{{\small{$J_1$}}} \psfrag{a}{{\small{$\alpha J_1$}}}
\psfrag{Jp}{{\small{$J_{\bot}$}}} \psfrag{J4}{{\small{$J_4$}}}
\epsfig{file=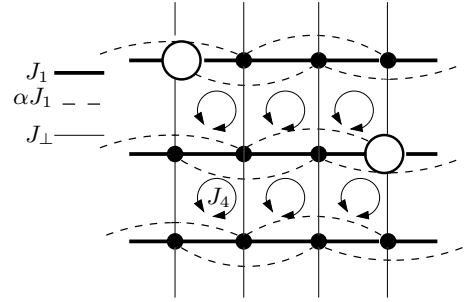,width=6cm}
\caption{Schematic picture
of the coupled chains model with nearest neighbor, next-nearest
neighbor, inter-chain and $4$-spins couplings $J_1$, $J_2=\alpha
J_1$, $J_{\bot}$, and $J_4$. Full (resp. open) circles stand for
spin-$\frac{1}{2}$ sites (resp. non-magnetic dopants).}
\label{fig:Lattice}
\end{center}
\end{figure}

The motivation of this Letter is to investigate by numerical
methods based on Exact Diagonalisations (ED) of finite chains (see
Ref.~\cite{Dobry98}) the role of various inter-chain magnetic
couplings on soliton confinement. Such couplings are expected to
become crucial in the case where magneto-elastic effects remain
small. Indeed, in that case, we argue that the formation of
magnetic moments in doped gapped frustrated spin chains is then
primarily controlled by 4-spin interactions.

Let us first consider a model of coupled frustrated
spin-$\frac{1}{2}$ antiferromagnetic (AF) chains, (see Fig.\ref{fig:Lattice})
\begin{eqnarray}
\label{hamil2D} H=\sum_{i,a} [ J_1(\vec S_{i,a}\cdot \vec
S_{i+1,a}
+ \alpha\vec S_{i,a}\cdot \vec S_{i+2,a}) \nonumber \\
+J_{\bot}\vec S_{i,a}\cdot \vec S_{i,a+1} ],
\end{eqnarray}
where $i$ is a lattice index along the chain of size $L$ and $a$
labels the $M$ chains ($L$ and $M$ chosen to be even). Periodic
boundary conditions will be assumed in {\it both directions}. In
the absence of inter-chain coupling ($J_{\bot}=0$) the behavior of
the un-doped chain is controlled by the frustration $\alpha$. From
$\alpha=0$ to $\alpha_c \simeq 0.241167$~\cite{Eggert96}, the
chain is in a critical regime with a gapless excitation spectrum.
Larger frustration stabilizes a spontaneously dimerised phase
characterized by a gap opening. At $\alpha=0.5$, the so-called
Majumdar-Ghosh~\cite{MG} (MG) point, the $2$-fold degenerate
Ground State (GS) consists in the product of singlets located
either on odd or on even bonds. Beyond the MG point, the
short-range correlations become incommensurate.

\begin{figure}
\begin{center}
\psfrag{A1}{{\small{$J_4=0.01$}}}
\psfrag{A2}{{\small{$J_4=0.05$}}} \psfrag{A3}{{\small{$J_4=0.1$}}}
\psfrag{A4}{{\small{$J_4=0.2$}}} \psfrag{A5}{$J_{\bot}$}
\psfrag{AF}{AF} \psfrag{SP}{SP} \psfrag{ac}{\small{$\alpha_c$}}
\psfrag{alp}{$\alpha$} \psfrag{L12}{\small{$L=12$}}
\psfrag{L16}{\small{$L=16$}} \psfrag{F S E}{{\small {F S E}}}
\epsfig{file=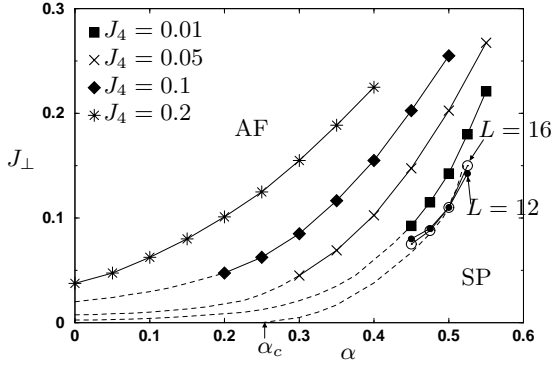,width=7cm} \caption{SP-AF phase diagram in
the ($\alpha,J_{\bot}$) plane from ED of chains of length up to
$16$ sites. Symbols correspond to different values of $J_4\ge 0$
as indicated on plot. Typically, FSE are smaller than the size of
the symbols. The computed transition lines are extended by {\it
tentative} transition lines (dashed lines) in the region where FSE
become large. At $J_4=0$ we have plotted a few points in the
vicinity of the MG point for $L=12$ and $L=16$.} \label{fig:PhDg2}
\end{center}
\end{figure}

The inter-chain AF coupling $J_{\bot}$ can stabilize a N\'eel
ordered phase under some conditions. Following
Schulz~\cite{Schulz96}, a mean-field (MF) treatment of the inter-chain
coupling is performed,
\begin{eqnarray}
\label{hamilQ1D.1imp} H_{\rm eff}(\alpha,J_{\bot})=\sum_{i,a}[\vec
S_{i,a}\cdot \vec S_{i+1,a} + \alpha\vec S_{i,a}\cdot
\vec S_{i+2,a} \nonumber\\
+h_{i,a}S_{i,a}^z-J_{\bot}\langle
S^{z}_{i,a}\rangle\langle S^{z}_{i,a+1}\rangle],
\end{eqnarray}
with
\begin{equation}
\label{Hi}
h_{i,a}=J_{\bot}(\langle S_{i,a+1}^z \rangle+\langle S_{i,a-1}^z \rangle)\, ,
\end{equation}
the local magnetic field computed self-consistently. In the
absence of dopant, we expect an homogeneous AF phase characterized
by a self-consistent staggered magnetization $\langle
S^{z}_{i,a}\rangle=(-1)^{i+a} m$ so that the system reduces to a
single chain under a staggered magnetic field $h_i=\pm 2(-1)^i
J_\perp m$. In the absence of frustration ($\alpha=0$), it was
shown that $m\sim \sqrt{J_{\bot}}$~\cite{Schulz96}. By solving the
self-consistency condition using ED of finite chains with up to
$16$ sites (supplemented by a finite size scaling analysis) the
transition line $J_{\bot}=J_{\bot}^c(\alpha)$ (see
Fig.~\ref{fig:PhDg2}) separating the  dimerised SP phase ($m=0$)
and the AF ordered phase (for which $m\neq 0$) has been obtained
in agreement with field theoretic approaches~\cite{Fukuyama96}.
Finite size effects are small in the gapped regime and especially
at the MG point ($J_{\bot}^c(\alpha=0.5)\simeq 0.11$ for all
sizes). Note also that numerical data suggest that the AF order
sets up at arbitrary small coupling when $\alpha < \alpha_c$ with
a clear finite size scaling $J_{\bot}^c(L)\propto 1/L$ at small
$\alpha$.

A non-magnetic dopant is
described here as an inert site decoupled from its neighbors.
Under doping the system becomes
non-homogeneous so that we define a local mean staggered
magnetization,
\begin{equation}
\label{MeStMg} {\mathcal{M}}^{\rm stag}_{i,a}=\frac{1}{4}
(-1)^{i+a}(2\langle S^{z}_{i,a} \rangle - \langle S^{z}_{i+1,a}
\rangle - \langle S^{z}_{i-1,a} \rangle).
\end{equation}
Following the method used in Ref.~\cite{Dobry98}, the MF equations
are solved self-consistently on finite $L\times M$ clusters and
lead to a non-uniform solution. At each step of the MF iteration
procedure, we use Lanczos ED techniques to treat {\it exactly}
(although independently) the $M$ {\it non-equivalent} finite
chains  and compute
$\langle S^{z}_{i,a}\rangle$ for the next iteration step
until convergence is eventually achieved.

\begin{figure}
\begin{center}
\psfrag{M1}{${\mathcal{M}}_{i,a}^{\rm stag}$} \psfrag{x}{$i$}
\psfrag{a}{{\small{$J_4=0$}}} \psfrag{b}{{\small{$J_4=0.01$}}}
\psfrag{c}{{\small{$J_4=0.08$}}}
\epsfig{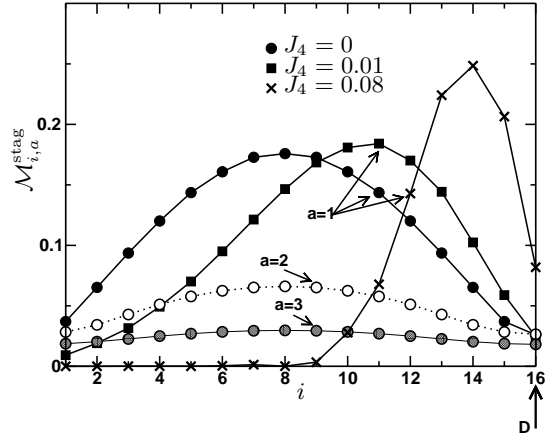} \caption{Local
magnetization ${\mathcal{M}}^{\rm stag}_{i,a}$ for $L\times
M$=$16\times 8$ coupled chains with one dopant D (shown by arrow)
located at $a=1$, $i=16$ in the dimerised phase ($\alpha=0.5$,
$J_{\bot}=0.1$). Circles correspond to $J_4=0$ (shown up to the
third neighboring chain of the doped one) and squares (crosses) to
$J_4=0.01$ ($J_4=0.08$). The coupling across the dopant has been
set to $0$ for convenience.} \label{fig:ConfinJ2p}
\end{center}
\end{figure}

We first consider the case of a single dopant. Whereas in the AF
phase the magnetization profile is weakly enhanced in the
vicinity of the dopant~\cite{Laukamp98,note_2D,Affleck2002}, in the SP gapped
phase the soliton remains de-confined as can be seen from
Fig.\ref{fig:ConfinJ2p}. Note that the inter-chain coupling
induces a ''polarization cloud'' with strong antiferromagnetic
correlations in the neighbor chains of the doped one.

The fact that our treatment of $J_\perp$ does not produce confinement is clear at the MG point where the dopant site separates
two regions with out-of-phase dimer coverage. In such a state, the
local magnetization $\left< S^{z}_{i}\right>$ identically
vanishes. Hence, we expect $J_{\bot}S_{i,a}^z \langle
S^{z}_{i,a\pm 1}\rangle$ to be irrelevant. On the contrary, terms
like  $\langle \vec S_{i}\cdot\vec S_{i+1}\rangle$ are $0$ or
$-3/4$ in the MG state and, hence, are relevant. Therefore, a
4-spin coupling $J_4(\vec S_{i,a}\cdot\vec S_{i+1,a})(\vec
S_{i+1,a+1}\cdot\vec S_{i,a+1})$ which originates from cyclic
exchange~\cite{Laeuchli2002} is of crucial
importance~\cite{note_J4}. As done for $J_{\bot}$ we have applied
a mean-field treatment to the $J_4$ coupling leading to a
(self-consistent) modulation of the intra-chain nearest neighbor
couplings of the Halmiltonian,
\begin{eqnarray}
\label{hamilQ1D.MF} H_{\rm eff}(\alpha,J_{\bot},J_4)
=\sum_{i,a}[(1+\delta J_{i,a})\vec S_{i,a}\cdot \vec S_{i+1,a}
\nonumber \\ + \alpha\vec S_{i,a}\cdot \vec
S_{i+2,a}+h_{i,a}S_{i,a}^z + {\rm constant}\, ,
\end{eqnarray}
with  $h_{i,a}$ given by Eq.(\ref{Hi}) and
\begin{equation}
\label{delti} \delta J_{i,a}=J_4\lbrace\langle \vec
S_{i,a+1}\cdot\vec S_{i+1,a+1} \rangle+\langle \vec
S_{i,a-1}\cdot\vec S_{i+1,a-1} \rangle\rbrace.
\end{equation}
As above, the numerical treatment again consists in a recursive
diagonalisation of the $M$ spin chains in the presence of the
(non-uniform) MF magnetic fields {\it and} exchange modulations
produced by its neighbors. Note that, as before, the effective
coupling between the chains appears only through the MF
self-consistency conditions so that, in practice, each chain
Hamiltonian can be Lanczos diagonalised independently {\it at each
iteration}.

Let us first consider the relative stability of the AF phase with
respect to the dimerised phase in the un-doped system with 4-spin
coupling. Since all chains become equivalent we have to deal with
an effective single dimerised frustrated chain problem in a
staggered magnetic field. If the chains are not dimerised, $\delta
J_{i,a}$ is constant and the exchange is just renormalised. On the
other hand, if the SP phase is stable, each chain displays the
same dimerised pattern when $J_4<0$ whereas dimer order is
staggered in the transverse direction when $J_4>0$. Note that,
apart from special features (see below), physical properties for
positive or negative $J_4$ are quite similar. Note also that the
dimerised GS would be $2^M$-fold degenerate when $J_4=0$ (each
chain is independently two-fold degenerate) while the degeneracy
is reduced to 2 when $J_4 \neq 0$. When $\alpha=J_{\bot}=0$ but
$J_4\neq 0$ each chain spontaneously dimerises and a gap opens up.
Consequently, a finite value of the AF inter-chain coupling is
necessary to drive the system into the AF ordered
phase~\cite{Matsumoto02}. The frustration $\alpha$ stabilizes
further the dimerised phase with respect to the AF one, the
critical $J_\perp(\alpha)$ increasing with increasing $\alpha$ as
seen from the phase diagram shown in Fig.~\ref{fig:PhDg2}. Note
that finite size effects (FSE) become large at small $J_4$ and
$\alpha$ and accurate extrapolations are not possible there.

Contrary to the previous case the spin-$\frac{1}{2}$ soliton
released in a dimerised state by doping experiences a confining
string which binds it to the dopant. Indeed, the $J_4$ interaction
stabilizes (if one assumes an infinite number of chains) a given
dimerisation pattern (over the two possible).
Fig.~\ref{fig:ConfinJ2p} shows the enhancement of the staggered
magnetization (\ref{MeStMg}) in the doped chain in the vicinity of
the dopant when $J_4 \neq 0$. Note that the dopant side where the
soliton is bound is imposed by the (arbitrary) sign of the bulk
dimerisation. A local spin-1/2 magnetization is then expected in
the vicinity of the dopant.

\begin{figure}
\begin{center}
\psfrag{12}{$L=12$} \psfrag{L16}{$L=16$}
\psfrag{S}{{\small{$\langle S_{i,1}^z \rangle$}}}
\psfrag{z}{{\small{$i$}}} \psfrag{J4}{$J_{4}$} \psfrag{xi}{$\xi$}
\psfrag{lo}{{\small{$2\xi$}}} \epsfig{file=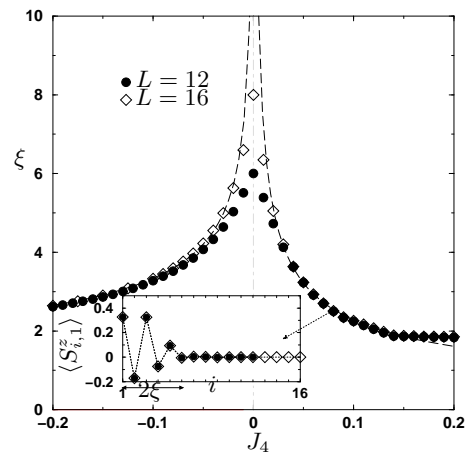,width=6cm}
\caption{ED data of the soliton average position  vs $J_4$
calculated for $\alpha=0.5$ and $J_{\bot}=0.1$. Different symbols
are used for $L\times M$ = $12\times 6$ and $16\times 8$ clusters.
The long-dashed line is a power-law fit (see text). Inset shows
the magnetization profile in the doped ($a=1$) chain at $J_4=0.08$, ie
$\xi \simeq 2.5$.} \label{fig:xi}
\end{center}
\end{figure}

In order to "measure" the strength of confinement, we define an
averaged soliton-dopant distance as $\xi=\sum_{i}i|\langle
S_i^z\rangle|/\sum_i |\langle S_i^z\rangle|$ so that $\xi=L/2$ in
the absence of confinement ($J_4=0$) and $\xi$ converges to a
finite value otherwise when $L\rightarrow\infty$. On
Fig.\ref{fig:xi}, $\xi$ is plotted versus $J_4$ for $2$ different
system lengths and $\alpha=0.5$ and $J_{\bot}=0.1$. FSE decrease
for increasing $J_4$.  Note that $\xi(J_4)\neq\xi(-J_4)$ and a
power law~\cite{Nakamura99} with different exponents $\eta$ is
expected when $J_4\rightarrow 0$. A fit gives $\eta\sim 0.33$ if
$J_4<0$ and $\eta\sim 0.50$ for $ J_4>0$ (Fig.\ref{fig:xi}). This
asymmetry can be understood from opposite renormalisations of
$J_1$ for different signs of $J_4$. Indeed, if $J_4<0$ then
$\delta J_{i,a}>0$ and the nearest neighbor MF exchange becomes
larger than the bare one. Opposite effects are induced by $J_4>0$.

\begin{figure}
\begin{center}
\psfrag{Jeff}{{{$|J^{\rm eff}(\Delta a,\Delta
i)|$}}} \psfrag{a}{\small{(a)}}
\psfrag{b}{\small{(b)}} \psfrag{c}{\small{(c)}}
\psfrag{A}{\small{(A)}} \psfrag{B}{\small{(B)}}
\psfrag{C}{\small{(C)}} \psfrag{d}{\small{(d)}}
\psfrag{i}{\small{(i)}} \psfrag{2i}{\small{(ii)}}
\psfrag{a0}{\small{$\Delta a=0$}} \psfrag{a1}{\small{$\Delta a=1$}}
\psfrag{a2}{\small{$\Delta a=2$}} \psfrag{Di}{$\Delta i$}
\epsfig{file=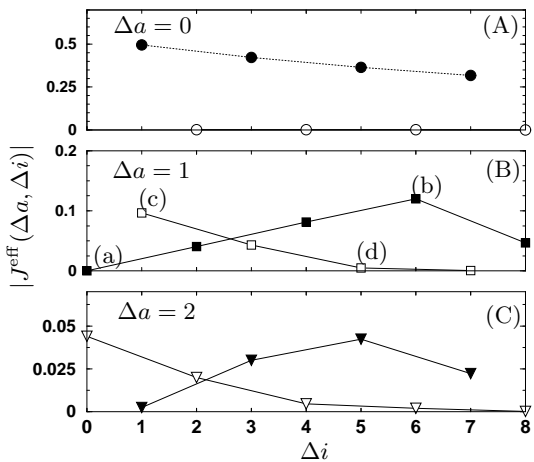,width=7cm} \caption{Magnitude of the
effective magnetic coupling between two impurities located either
on the same chain $(A)$ ($\Delta a=0$) or on different ones $(B)$
\& $(C)$ ($\Delta a=1,2$)  vs the dopant separation $\Delta i$ in
a system of size $L \times M=16 \times 8$ with $\alpha=0.5$,
$J_{\bot}=0.1$, and $J_4=0.08$. Closed (resp. open) symbols
correspond to AF (F) interactions.}

\label{fig:Jeff}
\end{center}
\end{figure}

We now turn to the investigation of the effective interaction
between dopants. We consider here a system of coupled chains with
two dopants (see Fig.~\ref{fig:Lattice}). Each dopant releases an
effective spin $\frac{1}{2}$ which is localized at a distance
$\sim \xi$ from it due to the confining potential set by $J_4$. We
define an effective pairwise interaction $J^{\rm eff}$ as the
energy difference of the $S=1$ and the $S=0$ GS. When $J^{\rm
eff}=E(S=1)-E(S=0)$ is positive (negative) the spin interaction is
AF (ferromagnetic). Let us first consider the case of two
dopants in the same chain. (i) When the two vacancies are on the
same sub-lattice the moments experience a very small ferromagnetic
$J^{\rm eff}<0$ as seen in Fig.~\ref{fig:Jeff}$(A)$ so that the
two effective spins $\frac{1}{2}$ are almost free. (ii) When the
two vacancies sit on different sub-lattices, $\Delta i$ is odd and
the effective coupling is AF with a magnitude close to the
singlet-triplet gap. Fig.~\ref{fig:Jeff}$(A)$ shows that the decay
of $J^{\rm eff}$ with distance is in fact very slow for such a
configuration. Physically, this result shows that a soliton and an
anti-soliton on the same chain and different sublattices tend to
recombine.

\begin{figure}
\begin{center}
\psfrag{i}{${i}$} \psfrag{M}{${\mathcal{M}}^{\rm stag}_{i,a}$}
\psfrag{a}{\small{(a)}} \psfrag{b}{\small{(b)}}
\psfrag{c}{\small{(c)}} \psfrag{d}{\small{(d)}}
\psfrag{d1}{\small{$D_1$}} \psfrag{d2}{\small{$D_2$}}
\epsfig{file=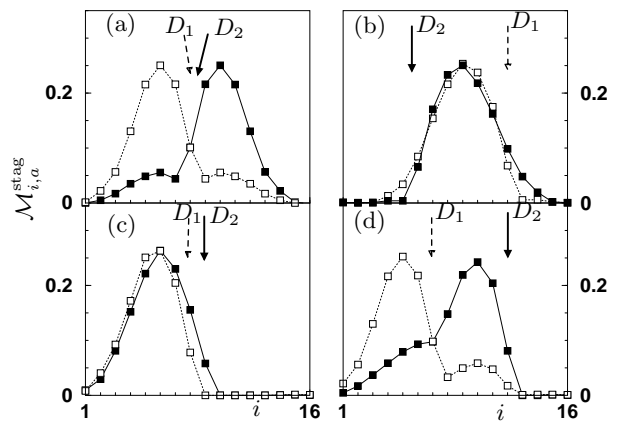,width=8cm} \caption{Local staggered
magnetization ${\mathcal{M}}^{\rm stag}_{i,a}$ for 2 dopants
($D_1$ \& $D_2$)  on
neighboring chains. The labels $(a)$-$(d)$ refer to the data
points shown on Fig.~\ref{fig:Jeff}$(B)$ corresponding to
different relative positions of the dopants. Arrows stand for the
dopant positions in the chain direction
and full (open) symbols correspond to the chain
doped with $D_2$ ($D_1$).} \label{fig:Ma2imp}
\end{center}
\end{figure}

The behavior of the pairwise interaction of two dopants located on
{\it different} chains ($\Delta a=1,2$) is shown on
Fig.~\ref{fig:Jeff}$(B)$ \& $(C)$ for $J_4>0$. When dopants are on
opposite sub-lattices the effective interaction is
antiferromagnetic. At small dopant separation $J^{\rm eff}(\Delta
i)$ increases with the dopant separation as the overlap between
the two AF clouds increases until $\Delta i \sim 2\xi$. For larger
separation, $J^{\rm eff}(\Delta i)$ decays rapidly. Note that the
released spin-$\frac{1}{2}$ solitons bind on the opposite right
and left sides of the dopants as imposed by the the bulk
dimerisation~\cite{note2}. If dopants are on the same sub-lattice,
solitons are located on the same side of the dopants~\cite{note3}
and the effective exchange $J^{\rm eff}(\Delta i)$ is
ferromagnetic and decays rapidly to become negligible when $\Delta i >
2\xi$. The behavior of the GS local staggered magnetization is
plotted on Fig.~\ref{fig:Ma2imp} for parameters corresponding to
$4$ typical behaviors. The key result here is the fact that the
effective pairwise interaction is {\it not} frustrating (because
of its sign alternation with distance) although frustration is
present in the microscopic underlying model. AF ordering is then
expected (at $T=0$) as seen for a related system of coupled
Spin-Peierls chains~\cite{Dobry98}.

We finish with a discussion on the relevance to materials.
Coexistence of AF ordering and SP order observed in Zinc-doped
CuGeO$_3$~\cite{Zn_CuGeO} can be understood as an ordering of the
induced $S=1/2$ moments in the dimerised
background~\cite{Dobry98}. We have shown that a 4-spin magnetic
inter-chain interaction, similarly to the coupling to a 2D (or 3D)
adiabatic lattice~\cite{Dobry98} but in contrast to a pairwise
inter-chain coupling, leads to the formation of local magnetic
moments under doping by inert atoms. Intermediate-range {\it
non-frustrated} spin exchange interactions between these moments
are expected to stabilize a low-temperature AF phase. Therefore,
cyclic exchange is expected to play a key role in quasi-1D
frustrated spin chains like LiV$_2$O$_5$~\cite{LiV2O5}.

We gratefully acknowledge stimulating discussions with I.~Affleck
and E.~Sorensen.\\
{\it Note added}.--Anadditional small confinement $\propto J_{\perp}^{2}/J_1$ was found in second order perturbation in $J_{\perp}$~\cite{Byrnes99}.


\begin{references}
\bibitem{note_J4} Our definition of $J_1$ includes the small
contribution from cyclic exchange. We have shown that the other
bilinear terms produced by cyclic exchange have negligible
effects.
\bibitem{note2} In that case, the solution is unique and
the sign of the bulk dimerisation is fixed by the local condition
of maximum overlap between the 2 solitonic clouds.
\bibitem{note3} Two equivalent symmetric solutions are obtained with
opposite signs of the bulk dimerisation.
\end{references}
\end{document}